\documentstyle[aps,eqsecnum,preprint,floats,epsf,epsfig]{revtex}
\input epsf.tex
\def\DESepsf(#1 width #2){\epsfxsize=#2 \epsfbox{#1}}
\begin{document}
\pagestyle{empty}                                      
\preprint{
\font\fortssbx=cmssbx10 scaled \magstep2
\hbox to \hsize{
\hbox{
            }
\hfill $
\vtop{
 \hbox{ }}$
}
}
\draft
\vfill
\title{Tree-Penguin Interference  \\
and \\
Tests for {$\cos\gamma < 0$} in Rare
$B\to PP$, $PV$ and $VV$ Decays}
\vfill
\author{
$^{1}$Wei-Shu Hou and $^{2,3}$Kwei-Chou Yang}
\address{
\rm $^1$Department of Physics, National Taiwan University, Taipei,
Taiwan 10764, R.O.C.\\
 \rm $^2$Institute of Physics, Academia
Sinica, Taipei, Taiwan 11529, R.O.C.\\
 \rm $^3$Department of Physics, Chung Yuan Christian University,
 Chung-li, Taiwan 32023, R.O.C.}

\date{\today}
%
%
\vfill
\maketitle
\begin{abstract}
Recent rare $B \to PP$, $PV$ decay data suggest that
factorization holds well if,
contrary to current fits, one has $\cos\gamma < 0$ where
$\gamma \equiv {\rm arg}(V_{ub}^*)$.
We update previous results with
light cone sum rule form factors,
which seem to work better.
We then discuss various $B\to VV$ modes
as well as the $K^{*} \eta$ modes.
Finding the pattern of $\rho^+\omega^0 < \rho^+\rho^0$,
$K^{*+}\rho^{-,0} > K^{*0}\rho^+$,
$K^{*+}\omega^0 > K^{*0}\omega^0$
and $K^{*+}\eta > K^{*0}\eta$ would strengthen the support
for $\cos\gamma < 0$.
The electroweak penguin enhances (suppresses) the $K^{*+}\rho^{0}$
($K^{*0}\rho^{0}$) rate by a factor of 2,
and finding $K^{*+}\rho^{0} \simeq K^{*+}\rho^{-}$
would be strong evidence for the electroweak penguin.
\end{abstract}
\pacs{PACS numbers:
}
\pagestyle{plain}

\section{Introduction}

Experimentally, a number of hadronic rare B decay modes have been
observed~\cite{jima,omegaK,gaofrank} in the last two years.
They may allow us access \cite{FM,DHHP,NR,HHY}
to unitarity angles of the Kobayashi-Maskawa (KM) matrix such as
$\gamma\ (\equiv {\rm arg} (V_{ub}^*)$ in standard phase convention),
by exploiting interference between tree and penguin amplitudes in these modes.
The presently observed decay processes can be catalogued into two classes.
The first class, e.g. $B\to \rho \pi$,
is dominated by tree (T) level $b\to u$ transitions,
but may have sizable penguin (P) contributions.
The second class, e.g. $B\to K\eta^\prime,\ K\pi,$
and the newly observed $K^{*+} \pi^-$ mode,
are penguin dominant processes which may have sizable T/P.

Two-body decays of B mesons are usually studied under the
factorization hypothesis.
Based on this hypothesis, the decay amplitude is given in
terms of a weak transition amplitude and the decay constant of a
factorized final state meson.
Nonfactorizable contributions are lumped into
the effective number of colors $N_{\rm eff}$
which may deviate from $N_c=3$.
The current fits of KM parameters
give $\gamma$ in the range of $60^\circ-70^\circ$~\cite{fit},
which heavily relies on the lower limit
$\Delta m_{B_s}>12.4~{\rm ps}^{-1}$ from combining LEP, CDF and SLD data.
With a little loosened limit
$\Delta m_{B_s}>10.2~{\rm ps}^{-1}$~\cite{PDG} at 95\% C.L.,
some room is allowed for negative ${\rm cos}~\gamma$.
If one adopts, however,
the currently favored $\gamma \simeq 60^\circ-70^\circ$,
it is difficult to explain present data such as
$K^+ \pi^-\sim K^0 \pi^+\sim K^+ \pi^0 \sim 1.5 \times 10^{-5}$,
$\pi^+ \pi^-< 0.84\times 10^{-5}$,
and the strength of the newly observed
$\rho^0\pi^+ \sim 1.5 \times 10^{-5}$
and $K^{*+} \pi^- \sim 2.2 \times 10^{-5}$\cite{gaofrank}.
All the data so far therefore seem to prefer $\cos\gamma < 0$ if
factorization holds~\cite{HHY}, except the size of
$K^+\omega^0 \sim 1.5 \times 10^{-5}$ \cite{omegaK}
which cannot be explained by factorization
[See Note Added.].
However, all modes with branching ratios (Br) of order $10^{-5}$ or more
will likely be updated or measured soon by CLEO and the B factories.
It is thus of interest to explore any additional modes
that can shed further light on $\gamma$.
In this paper we extend Ref. \cite{HHY} and study
additional channels \cite{TC} for which the $\gamma$ range can be probed.

We update the $B\to PP$ and $PV$ modes
($P$, $V$ stand for pseudoscalar and vector mesons)
with form factors from light-cone (LC) sum rules \cite{LC},
which seem to give a better fit to data
than using Bauer-Stech-Wirbel (BSW) form factors \cite{BSW}.
We find further that some $VV$ modes
and the $K^*\eta$ modes are promising.
Processes that are basically pure T (e.g.
$\rho^+\rho^0$) or pure P (e.g. $K^{(*)}\phi$) depend only
weakly on $\gamma$, and thus offer direct tests of factorization.
If large CP asymmetries ($a_{CP}$) are observed in the
$K^{(*)}\phi$ modes, it could be a signal for new physics.
The paper is organized as follows.
In Sec.~\ref{sec:theory} a
brief review of the theoretical framework is given.
We then sketch how sensitivity to $\gamma$ angle emerges.
In Sec.~\ref{sec:data} we
discuss in detail the hints of negative
${\rm cos}\gamma$ from existing data.
We show that the form factors from LC sum rules are preferred by data.
Adopting LC sum rule form factors, in Sec.~\ref{sec:VV} we
study the $VV$ modes as well as some other modes
that can offer further tests for $\cos\gamma < 0$
or the factorization hypothesis.
Finally, the discussion and conclusion are presented in Sec.~\ref{dc}.

\vskip 0.7cm
\section{Theoretical framework}\label{sec:theory}

The standard starting point is the
effective $\Delta B=1$ weak Hamiltonian
\begin{eqnarray}
{\cal H}_{\rm eff}&=& {G_F\over\sqrt{2}}
\Bigg\{ V_{uq}^*V_{ub}\Big[c_1(\mu)O_1^u(\mu)+c_2(\mu)O_2^u(\mu)\Big]
       +V_{cq}^*V_{cb}\Big[c_1(\mu)O_1^c(\mu)+c_2(\mu)O_2^c(\mu)\Big]
         \nonumber \\
&& \ \ \ \ \
       -V_{tq}^*V_{tb}\sum^{10}_{i=3}c_i(\mu)O_i(\mu)\Bigg\}+{\rm h.c.},
\end{eqnarray}
where $q=d,s$, and
\begin{eqnarray}
&& O_1^u = (\bar qu)_{V-A}(\bar ub)_{V-A}, \qquad\qquad\qquad\quad~~
   O_2^u = (\bar q_\beta u_\alpha)_{V-A}(\bar u_\alpha b_\beta)_{V-A},
    \nonumber \\
&& O_1^c = (\bar qc)_{V-A}(\bar cb)_{V-A}, \qquad\qquad\qquad\quad~~~
   O_2^c = (\bar q_\beta c_\alpha)_{V-A}(\bar c_\alpha b_\beta)_{V-A},
    \nonumber \\
&& O_{3(5)}=\sum_{q'}(\bar q'q')_{{V-A}({V+A})}(\bar qb)_{V-A}, \qquad \quad
   O_{4(6)}=\sum_{q'}(\bar q'_\beta q'_\alpha)_{{V-A}({V+A})}
            (\bar q_\alpha b_\beta)_{V-A}, \\
&& O_{7(9)}={3\over 2}\sum_{q'}e_{q'}(\bar q'q')_{{V+A}({V-A})}
            (\bar qb)_{V-A}, \quad
   O_{8(10)}={3\over 2}\sum_{q'}e_{q'}(\bar q'_\beta q'_\alpha)_{{V+A}({V-A})}
             (\bar q_\alpha b_\beta)_{V-A},   \nonumber
\end{eqnarray}
with $O_{3-6}$, $O_{7-10}$ the QCD, electroweak penguin operators
and $(\bar q_1q_2)_{_{V\pm A}}\equiv\bar q_1\gamma_\mu(1\pm \gamma_5)q_2$.
The decay amplitude is computed by evaluating the hadronic matrix elements
of ${\cal H}_{\rm eff}$, i.e.
\begin{eqnarray}
        c_i(\mu) \langle O_i(\mu)\rangle
 =      c_i(\mu) g_{ij}(\mu)\langle O_{j}\rangle_{\rm fac}
 \equiv c_j^{\rm eff}\langle O_j\rangle_{\rm fac},
\end{eqnarray}
where the $\mu$-dep. of $\langle O_i(\mu)\rangle$
has been taken out through the matrix $g_{ij}(\mu)$
which cancels the $\mu$-dep. of $c_i(\mu)$
to give $c_j^{\rm eff}$,
which should not depend on the theoretical scale parameter $\mu$.
The matrix elements $\langle O_j\rangle_{\rm fac}$
are evaluated at the factorization scale $\mu_f$
by equating it to products of matrix elements of quark bilinears,
the evaluation of which is done by form factor models.
It can be shown that the $c_i^{\rm eff}$s are $\mu$, scheme and
gauge independent~\cite{CLY},
but it should be at the same scale $\mu_f$
where one evaluates $\langle O_j\rangle_{\rm fac}$.
Whether, or how, factorization actually works, however,
is not well understood.

The decay amplitudes derived from the factorization approach
are given in terms of effective parameters $a_i^{\rm eff}$,
where $a_{2j}^{\rm eff} = {c}_{2j}^{\rm eff}+{1\over N_c}{c}_{2j-1}^{\rm eff}$
and $a_{2j-1}^{\rm eff} = {c}_{2j-1}^{\rm eff}+{1\over N_c}{c}^{\rm eff}_{2j}$
$(j=1,\cdots,5)$.
In what follows, we adopt the values of $a_i^{\rm eff}$
given in Ref.~\cite{CCTY} which are evaluated at $\mu_f = m_b$,
use $N_{c}=3$,
and ignore final state interactions (FSI).
Since the presently observed modes are largely color allowed,
most results here are insensitive to $N_{\rm eff} \neq N_c$.
The influence of $N_{\rm eff}\not= 3$ will be briefly discussed.
For detailed formulas we refer to Refs.~\cite{CCTY} and \cite{ali}.
We will take $q^2 = m_b^2/2$ \cite{GH} in penguin coefficients
to generate favorable absorptive parts.
Smaller $q^2$ values would lead to much smaller $a_{CP}$s.
Thus, the $CP$ asymmetries that we present are for sake of
showing the trend only.
As an indication of possible sensitivity to factorization scale $\mu_f$,
we list $a_i^{\rm eff}$ for $\mu_f = m_b$ and $m_b/2$ in Table I.

\vskip 0.1cm
\begin{table}[ht]
\vspace{1cm}
 {\small Table I.~Values for $a_i^{\rm eff}$ for $b\to s\bar qq$ processes
 for $N_c = 3$,
 evaluated at $\mu_f = m_b$ (first row) and $m_b/2$,
 where $\mu_f$ is the ``factorization scale"
 ($a_{3{\rm -}10}^{\rm eff}$ are in units of $10^{-4}$).
 We take $q^2 = m_b^2/2$ in determining the imaginary parts.}
\begin{center}
\begin{tabular}{ c c c c c c c c c c}
 $a_1$  & $a_2$ & $a_3$ & $a_4$ & $a_5$  & $a_6$ & $a_7$ & $a_8$ &
 $a_9$  & $a_{10}$  \\ \hline
 $1.046$   & $+0.024$ & $72$  & $-383 - 121i$ & $-27$  & $-435 - 121i$ &
 $-0.9 - 2.7i$ & $3.3 - 0.9i$ & $-93.9 - 2.7i$ & $0.3 - 0.9i$ \\
 $1.059$   & $-0.048$ & $96$  & $-396 - 120i$ & $-54$  & $-514 - 120i$ &
 $-0.5 - 2.7i$ & $4.0 - 0.9i$ & $-93.2 - 2.7i$ & $3.6 - 0.9i$ \\
\end{tabular}
\end{center}
\end{table}

\begin{figure}[htb]
\vspace{1cm}
   \centerline{\epsfig{figure=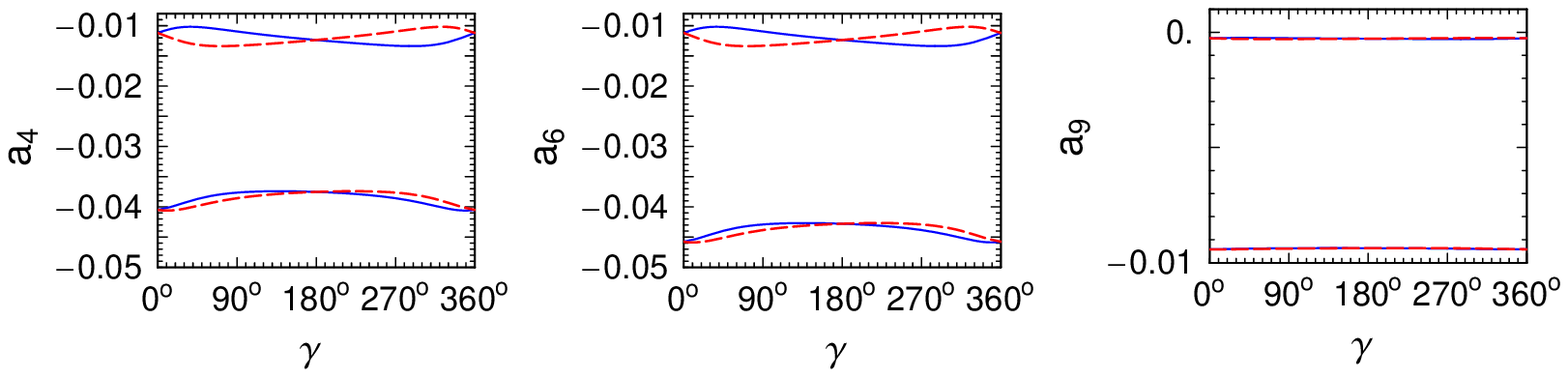,width=17cm}}
\vspace{0.3cm}
 \caption {Penguin coefficients $a_4,a_6$ and $a_9$ vs. $\gamma$,
 where the solid (dashed) curve are for $b\to d \bar qq$
 $(\bar b\to \bar d \bar qq)$ and the upper (lower) curves corresponds to
 ${\rm Re}(a_i)$ (${\rm Im}(a_i)$). \label{fig:ai}} \vspace{0.5cm}
\end{figure}

As we are interested in studying $\gamma$ dependence of
decay amplitudes, it is important to check the
$\gamma$ dependence of short distance coefficients.
Although the $a_i$'s for $b\to s$ penguins are
basically $\gamma$-independent because
$V_{us}^*V_{ub}$ is much smaller than
$V_{ts}^*V_{tb} \simeq -V_{cs}^*V_{cb}$,
it is not the case for $b\to d$ penguins
since all three KM factors are on same footing in
$V_{ud}^*V_{ub} + V_{cd}^*V_{cb} + V_{td}^*V_{tb} = 0$.
Thus, for $b\to d$ penguins,
$a_{3-10}$ will also exhibit $\gamma$ dependence.
In Fig.~\ref{fig:ai} we show the $\gamma$
dependence of $a_4, a_6$, and $a_9$ for both $b\to d\bar q q$
and $\bar b\to \bar d  \bar q q$. These are the
dominant gluonic and electroweak penguin coefficients.
We see that for $\gamma = 50^\circ - 150^\circ$,
Re\,$a_4$ and Re\,$a_6$ are
within 3\% of $-0.0383$ and $-0.0437$, respectively,
while Re\,$a_9$ is constant.
These values are basically the same as $b\to s$ penguins.
Variations of Im$(a_{4,6})$ are more sizable but they are
less significant than Re$(a_{4,6})$ in
contributing to average rates.
Thus, given the present experimental uncertainties
as well as underlying uncertainties associated with the
factorization assumption, to first approximation the
$\gamma$-dependence of $b\to d$ penguin coefficients
can be safely ignored.
In the following numeric results, however,
$\gamma$-dep. of $b\to d$ penguin coefficients
have been taken into account.

Let us see how tree-penguin interference gives us
a bearing on $\cos\gamma$.
Using the standard phase convention \cite{PDG} of putting $CP$ phase in
$V_{ub} = \vert V_{ub} \vert \; e^{-i\gamma}$,
the tree amplitudes ($O_1$ and $O_2$) for
$b\to u\bar ud$ and $b\to u\bar us$ processes have the KM factors
\begin{eqnarray}
 V_{ud}^*V_{ub} \cong \vert V_{ub} \vert e^{-i\gamma}
\ \ \ \ {\rm and}\ \ \ \
 V_{us}^*V_{ub} \cong \lambda \vert V_{ub} \vert e^{-i\gamma},
\end{eqnarray}
respectively, where $\lambda \equiv \vert V_{us} \vert \cong 0.22$.
The penguin amplitudes ($O_{3-10}$),
on the other hand, are governed by the KM factors
\begin{eqnarray}
 V_{td}^*V_{tb} & = & - \left(V_{cd}^*V_{cb} + V_{ud}^*V_{ub}\right)
                \cong + \left(\lambda \vert V_{cb} \vert
                            - \vert V_{ub} \vert e^{-i\gamma}\right),
 \\
 V_{ts}^*V_{tb} & = & - \left(V_{cs}^*V_{cb} + V_{us}^*V_{ub}\right)
                \cong - \left(\vert V_{cb} \vert
                        + \lambda \vert V_{ub} \vert e^{-i\gamma}\right)
                \cong - \vert V_{cb} \vert,
\end{eqnarray}
where KM unitarity, implicit in Eq. (2.1), has been used,
and the last step for $V_{ts}^*V_{tb} \cong - \vert V_{cb} \vert$
is accurate to less than 2\%.
Since $\vert V_{ub} / V_{cb} \vert \simeq 0.08$, one finds
$\lambda - \vert V_{ub} / V_{cb} \vert \; \cos\gamma > 0$ always
hence the real parts of $V_{td}^*V_{tb}$ and $V_{ts}^*V_{tb}$
are opposite in sign.
Thus, {\it not only T-P interference for
$b\to u\bar ud$ and $b\to u\bar us$ processes
depend on the sign of $\cos\gamma$,
the interference effect is opposite between the two type of processes},
e.g. when constructive in $K^+\pi^{-,0}$ for $\cos\gamma < 0$,
it is destructive in $\pi^+\pi^-$,
which is precisely what is needed to explain data.

Such phenomena are of fundamental nature,
and offer a window on the phase angle $\gamma$,
but it can be obscured by long distance effects
such as $\pi^+\pi^- \to \pi^0\pi^0$ \cite{HHY,CCTY} rescattering.
However, the nonobservation of $B\to K\bar K$ and $\pi^0\pi^0$ modes
\cite{jima,omegaK,gaofrank} suggest that FSI rescattering effects
are not sizable,
except for the case of $B^+\to K^+\omega^0 \sim 1.5\times 10^{-5}$.
We shall await experimental confirmation of the latter [See Note Added.]
and note in the mean time that
factorizaton is more likely to work in
the $N_C$ insensitive modes such as the ones studied here.
We note in passing that some recent work in
applications of perturbative QCD to B decays
are beginning to reveal how factorization works \cite{factorize}.


\section{Comparison of $B \to PP$, $PV$ Modes with Data}\label{sec:data}

It was $B \to PP$, $PV$ data that {\it inspired} the
observation that factorization does work and
hinted at $\cos\gamma < 0$ in Nature.
The starting point was the $K\pi$ modes.
Ignoring the electroweak penguin (EWP), one typically expects $K^+
\pi^0/K^+ \pi^-\approx (1/\sqrt{2})^2$, where the factor of
$1/\sqrt{2}$ comes from the $\pi^0$ isospin wave function,
and the ratio is almost independent of $\gamma$.
The data, however, suggest that $K^+\pi^0$ is
as large as $K^+ \pi^-$ \cite{jima,gaofrank},
which imply that EWP may be important~\cite{DHHP,NR}.
Choosing larger $m_s$ to suppress strong penguin $a_6$ contribution,
and {\it $\gamma$ in the range of $90^\circ-130^\circ$ to
enhance $K^+\pi^{-}$ and $K^+\pi^{0}$} with respect to $K^0\pi^{+}$,
it was shown~\cite{DHHP} that the three observed $K\pi$ modes can be
suitably close to each other and the data are thus accommodated.

The $\pi^+\pi^-$ mode then presents a challenge.
It is color allowed and should be $T$-dominant,
and easier to see experimentally than the recently measured
$B^+\to \rho^0\pi^+$ and $B^0\to \rho^\pm\pi^\mp$ modes \cite{gaofrank}.
However, it is not yet observed [See Note Added.].
Without resorting to a small $N_{\rm eff}\sim 1$
or large final state rescattering phases,
it was pointed out that suppression of the $\pi^+\pi^-$ mode can be
elegantly achieved if $\cos\gamma < 0$, which would enhance the
$\rho^0\pi^+$ mode (and even more so if $m_u + m_d$ is on
the lighter side) and suppress $\rho^\pm\pi^\mp$ \cite{HHY}.
If the $A_0^{B\to \rho}(q^2 = m_{\pi}^2)$ form factor is larger than in BSW
model~\cite{BSW}, it could further help explain the strength of
$\rho^0\pi^+ \sim 1.5 \times 10^{-5}$ and the smallness of the
ratio $\rho^\pm\pi^\mp/\rho^0\pi^+ = 2.3 \pm 1.3$ \cite{gaofrank}.

The newly measured $K^{*+} \pi^-$ mode is also color allowed and
insensitive to $N_{\rm eff}$,
while the $F_1^{B\to \pi}(m_{K^*}^2)$ form factor
is constrained by $B\to K\pi$, $K^+\phi^0$, $\pi\pi$ and the
semileptonic $B\to \pi(\rho)l\nu$ data. The factorization approach
gives too low a value of $K^{*+}\pi^- < 0.7\times 10^{-5}$
\cite{HHY} for $\gamma \sim 60^\circ - 70^\circ$. Choosing a
larger $\gamma$ such as $\sim 120^\circ$, however, $K^{*+} \pi^-$
can easily reach $1.2\times 10^{-5}$ or more \cite{HHY} and
becomes more consistent with data.

The above observations are largely insensitive to $N_{\rm eff}$.
In Ref. \cite{HHY} BSW form factors were used.
In fact, the form factors from light-cone sum rules~\cite{LC}
seem to give a better fit to $B\to PP$, $PV$ data,
since the $A_0$ form factor is larger
while $F_{0,1}$ form factors are slightly lower than in BSW model.
We list the relevant form factor values at zero momentum transfer
for both BSW model and LC sum rules in Table~II.
The $q^2$ dependence of the LC sum rule
results can be referred to~\cite{LC}.
Note that hadronic charmless $B\to PP$ and $VP$ are
insensitive to the $q^2$ dependence of form factors
because of the smallness of $q^2$ in the factorization approach.
However, if $F_1^{B\to P}(q^2)$ has dipole $q^2$ dependence,
the $K^{*+}\pi^-$ rate can be enhanced by 12\%
because $q^2 = m_{K^*}^2$ is no longer negligible.

\vskip 0.1cm
\begin{table}[ht]
\vspace{1cm}
 {\small Table II.~Form factors at zero momentum
transfer in the BSW model \cite{BSW} and in the LC sum
rule calculations~\cite{LC}. The values given in the square
brackets are obtained in the LC sum rule analysis.}
\begin{center}
\begin{tabular}{ l l c c c c }
 Decay  & $F_1=F_0$ & $V$ & $A_1$ & $A_2$ & $A_0$ \\ \hline
 $B\to\pi$ & 0.333~[0.305] & & & &  \\
 $B\to K$ & 0.379~[0.341]& & & & \\
 $B\to\rho$ & & 0.329~[0.338] & 0.283~[0.261] &
 0.283~[0.223] & 0.281~[0.372] \\
 $B\to K^*$ & & 0.369~[0.458] & 0.328~[0.337] & 0.331~[0.203]
 & 0.321~[0.470] \\
\end{tabular}
\end{center}
\end{table}

At this point we caution that form factor models typically
do not have good reference to the factorization scale $\mu_f$
that enters $a_i^{\rm eff}$.
Thus, until one has a more complete model of
how factorization works, one should bear in mind the uncertainties
in $a_i^{\rm eff}$ that may follow from changing $\mu_f = m_b$ to $m_b/2$,
as reflected in Table I.
In the complete theory, there should again be no $\mu_f$ dependence.
We note that some progress has been made recently
in providing a QCD basis for why and how
factorization works \cite{factorize}.

\begin{figure}[htb]
\centerline{\epsfig{figure=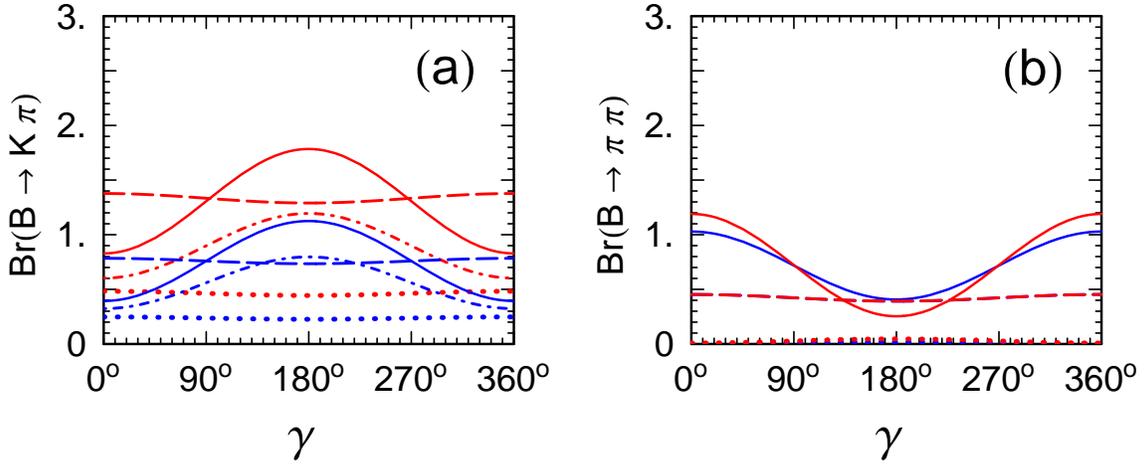,width=15cm} }
\smallskip
\caption {(a) Solid, dash, dotdash and dots for $B\to K^+\pi^-$,
$K^0\pi^+$, $K^+\pi^0$ and $K^0\pi^0$, for $m_s = $ 105 (upper
curves) and 200 MeV; (b) solid, dash and dots for $B\to
\pi^+\pi^-$, $\pi^+\pi^0$ and $\pi^0\pi^0$ for $m_d = 2m_u= $ 3
and 6.4 MeV, where the lower (upper) curve at $\gamma = 180^\circ$
for $\pi^+\pi^-$ ($\pi^0\pi^0$) is for lower $m_u+m_d$. In all
figures $\vert V_{ub}/V_{cb}\vert = 0.08$, $Br$s are in units of
$10^{-5}$, and light cone sum rule form factors are used.}
\label{fig:PP}
\end{figure}

\begin{figure}[htb]
\vspace{1cm}
 \centerline{\epsfig{figure=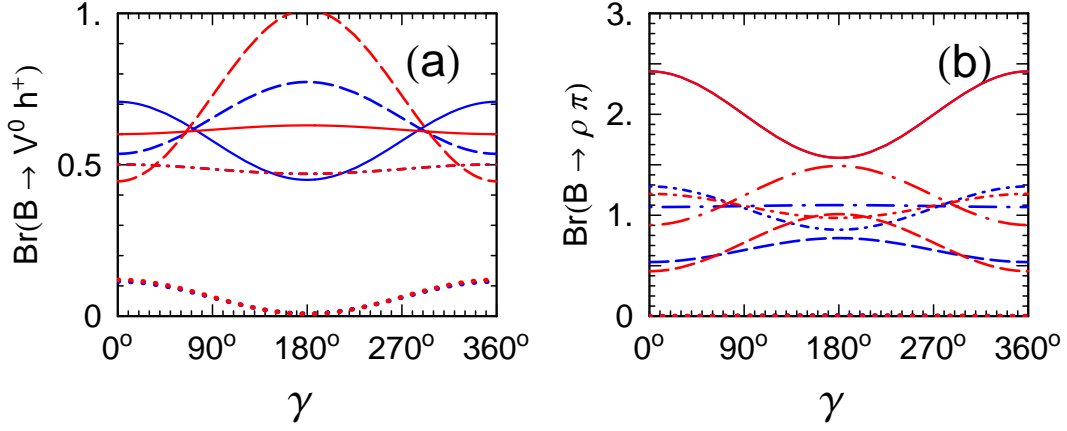, width=14cm} }
\smallskip
\caption { For $m_d = 2m_u= $ 3 and 6.4 MeV, (a) solid, dash,
dotdash and dots for $\omega^0\pi^+$, $\rho^0\pi^+$, $\phi^0 K^+$
and $\omega^0 K^+$; (b) solid, short-dotdash, long-dotdash, dash
and dots for $B\to \rho^+ \pi^-$, $\rho^+\pi^0$, $\rho^-\pi^+$,
$\rho^0 \pi^+$ and $\rho^0\pi^0$. The upper curves at $\gamma =
180^\circ$ are for lower $m_d$ and $m_u$.
} \label{fig:VP}
\end{figure}

\begin{figure}[htb]
\centerline{\epsfig{figure=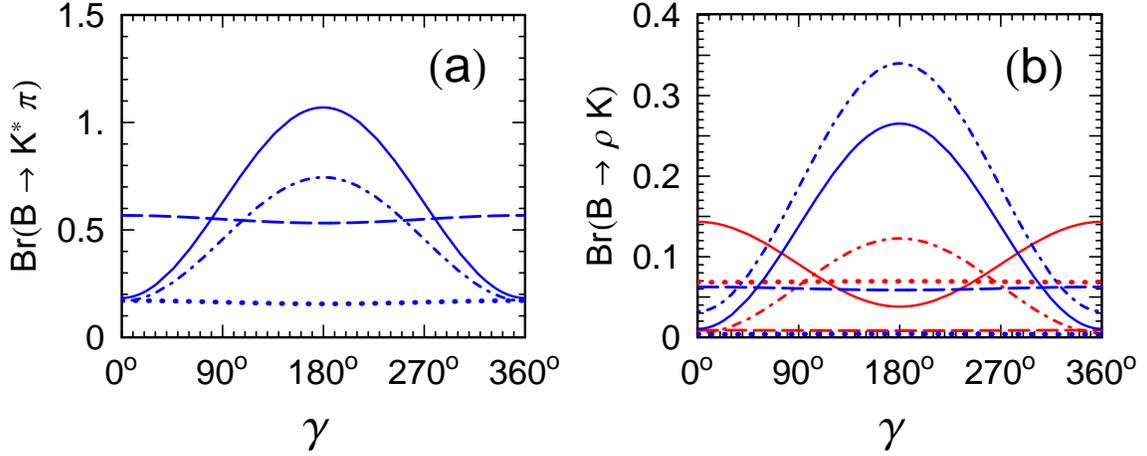, width=15cm}}
\smallskip
\caption { (a) Solid, dash, dotdash and dots for $B\to K^{*+}\pi^-
$, $K^{*0}\pi^+$, $K^{*+}\pi^0$ and $K^{*0}\pi^0 $, which are
insensitive to $m_s$. (b) Solid, dash, dotdash and dots for
$\rho^- K^{+}$, $\rho^+ K^{0}$, $\rho^0  K^{+}$ and $\rho^0
K^{0}$, for $m_s =$ 105 and 200 MeV. The upper (lower) curves for
$\rho K^0$ ($\rho K^+$) at $\gamma = 180^\circ$ are for lower
$m_s$. } \label{fig:rhoK}
\end{figure}

The results using BSW form factors have been given in \cite{HHY}.
Here, for comparison we use LC sum rule (LCSR) form factors
and plot the results versus $\gamma$ in
Figs.~\ref{fig:PP}, \ref{fig:VP} and \ref{fig:rhoK}.
The $K\pi$ and $\pi\pi$ modes fit data rather well,
except $K^+\pi^- > K^+\pi^0$ is expected if one picks $m_s \sim 100$ MeV.
As emphasized in~\cite{HHY},
a larger value of $A_0^{B\rho}$
(which is realized in the LCSR approach),
would pull up the $\rho^0\pi^+$ and $\omega^0 \pi^+$ rates.
Having $\rho^0\pi^+ > \omega^0 \pi^+$ which is hinted by data
would still prefer $\gamma \gtrsim 90^\circ$.
Because of a lower $F_1^{BK}$,
the Br of $\phi^0 K^+$ drops to $0.5\times 10^{-5}$,
which again fits better the experimental upper limit of
$0.59\times 10^{-5}$ \cite{gaofrank}.
The $\rho^\pm \pi^\mp$ rate is now lower because
$B^0\to \rho^+ \pi^-$ amplitude depends on $F_1^{B\pi}$ only,
while $B^0\to \rho^- \pi^+$ is enhanced by a
larger $A_0^{B\rho}$ analogous to $\rho^0\pi^+$.
For $\gamma = 120^\circ - 150^\circ$ and lighter $m_d + m_u$,
$\rho^\pm \pi^\mp \sim 3\times 10^{-5}$
and $\rho^0\pi^+$, $K^{*+}\pi^- \sim 1\times 10^{-5}$.
These values are lower than but within range of
recent CLEO observations \cite{gaofrank}.

Because the form factors from LC sum rule
calculations fit data better,
we adopt the LCSR form factors in subsequent
analysis of further modes.

\section{Analysis of $\gamma$-dependence of Further
Modes}\label{sec:VV}

\subsection{$B\to \rho\rho$ and $\rho\omega$ Modes}

\begin{figure}[htb]
\vspace{1cm}
   \leftline{\epsfig{figure=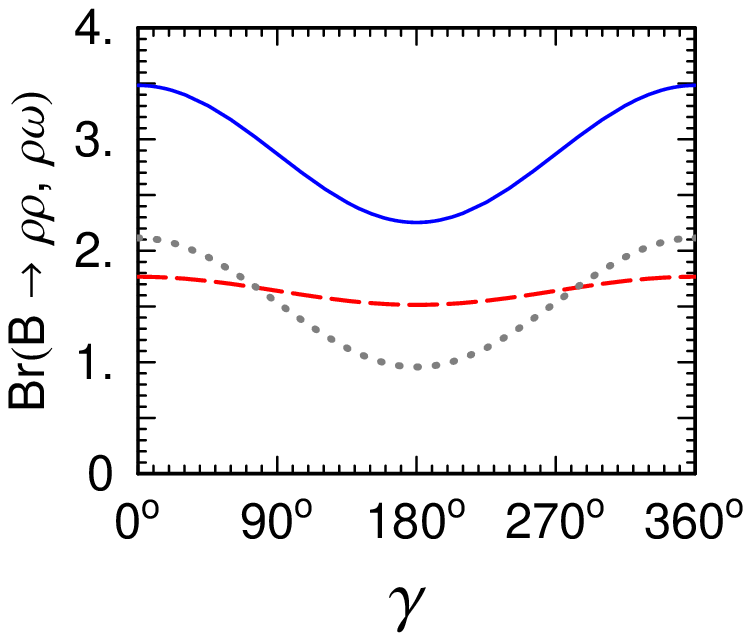,width=7.7cm,height=6cm}
    ~~~~~~~\epsfig{figure=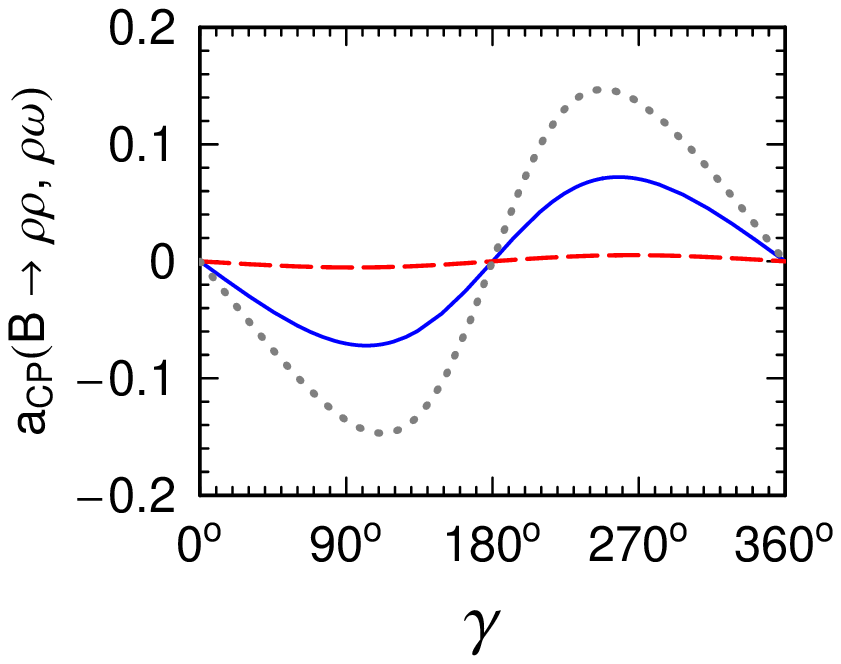,width=7.7cm,height=6cm}}
\vspace{0.8cm}
 \caption {$Br$s and $a_{CP}$s vs. $\gamma$
 where solid, dash and dots are for
 $\rho^+\rho^-$, $\rho^+\rho^0$ and $\rho^+\omega^0$, respectively.
\label{fig:rhorho}} \vspace{0.5cm}
\end{figure}

$B\to VV$ amplitudes are
independent of light quark masses.
The modes $\rho^+ \rho^-, \rho^+ \rho^0$, and
$\rho^+ \omega^0$ are all of order $10^{-5}$
with $\rho^+ \rho^-$ being the largest.
One expects $\rho^+\rho^-/\rho^+\omega^0\approx (1/\sqrt{2})^2$
where $1/\sqrt{2}$ comes from the $\omega^0$ isospin wave function.
The $\gamma$-dependence of $\rho^+\rho^-$ and $\rho^+ \omega^0$ rates
is dominated by the interference term
$\propto$ Re($V_{ud}^* V_{ub}a_1$) $\times$ Re$(V_{td}^* V_{tb}\, a_4)$.
In contrast, the $\rho^+ \rho^0$ mode is far less sensitive to
$\gamma$ since $a_4$ is replaced by $3a_9/2$
where $a_9$ is $\sim 4$ times smaller than $a_4$.
In any case, all three modes get suppressed for
$\cos\gamma < 0$, as shown in Fig.~\ref{fig:rhorho}.
For the currently favored value of
$\gamma\sim 60^\circ-70^\circ$ \cite{fit},
one expects
$\rho^+ \rho^- : \rho^+ \rho^0 : \rho^+ \omega^0 \simeq 3.1 : 1.7 : 1.7$
(roughly $\times 10^{-5}$),
but if $\gamma$ is larger than $90^\circ$,
say $\sim 120^\circ$, it becomes $2.5:1.6:1.2$,
reaching down to $2.3:1.5:1.0$ at $\gamma \sim 180^\circ$.
Thus, finding $\rho^+\omega^0 < \rho^+\rho^0$
would support $\cos\gamma < 0$,
similar to $\omega^0\pi^+ < \rho^0\pi^+$.
The branching ratios imply that these modes could be observed soon.
However, $\rho^+ \rho^- \to \pi^+\pi^0\pi^-\pi^0$
has two $\pi^0$'s in the final state
and would be harder to detect than the other two modes,
while $\rho^+\omega^0$ is expected to have the least background.

To study model dependence, we have also used form factor values
from BSW model~\cite{BSW} as input parameters. We find that the
ratios do not change much, but the overall scale can become
smaller by 40\%.

The $a_{CP}$s are dominated by
Im($V_{ud}^* V_{ub})a_1\, $Re$(V_{td}^* V_{tb})$ times
Im($a_4)$, $2\,$Im($a_4)$
and Im($3a_9/2)$ terms for $\rho^+\rho^-$,
$\rho^+ \omega^0$ and $\rho^+\rho^0$, respectively.
As seen from Fig.~\ref{fig:rhorho},
the $a_{CP}$s for $\rho^+\rho^-$, $\rho^+ \omega^0$
could be as large as $-7\%$, $-16\%$, respectively,
for $\gamma=90^\circ-130^\circ$,
while for $\rho^+ \rho^0$ it is very small
since the strong P contribution is forbidden by isospin.
The $a_{CP}$s are smaller for $\gamma \sim 60^\circ - 70^\circ$.

\subsection{$B\to K^*\rho$ Modes and the Electroweak Penguin}

Tree--penguin interference for
$K^{*+}\rho$ and $\rho^+\rho$ modes
differ in sign because the KM factors
${\rm Re}(V_{td}^* V_{tb})\cong -A\lambda^3(1-\rho)$ and
${\rm Re}(V_{ts}^* V_{tb})\cong -A\lambda^2$ have opposite sign,
quite analogous to the case of $K^+ \pi^{-,0}$ vs. $\pi^+\pi^-$ \cite{HHY}.
Thus, while $\rho^{+}\rho^-$ and $\rho^+\omega^0$ are
suppressed for $\cos\gamma  < 0$,
$K^{*+} \rho$ modes are enhanced.
Furthermore, the impact of EWP on $K^{*}\rho^0$ modes is
more prominent than on the $K\pi^0$ \cite{DHHP}
and $K^*\pi^0$ \cite{HHY} modes which have similar amplitude structure.

Let us show how the latter comes about.
For $K^{+} \pi^0/ K^{+} \pi^-$, we have
\begin{eqnarray}
\frac{K^{+} \pi^0}{K^{+} \pi^-}
\approx {1\over 2}
\left\vert 1 + r_0 \,
{\lambda \vert\frac{V_{ub}}{V_{cb}}\vert e^{-i\gamma}\, a_2
           + {3\over 2}a_9
\over
        \lambda \vert\frac{V_{ub}}{V_{cb}}\vert e^{-i\gamma}\, a_1
       + a_4 +a_6 R_4}\right\vert^2
\approx
 \left\{
      \begin{array}{ll}
       0.65, & m_s=105\ {\rm MeV} \\
       {\cal O}(1), & m_s\ {\rm large}.
      \end{array} \right.
\end{eqnarray}
where the factor of 1/2 is from the $\pi^0$ isospin wave function,
$r_0 = F_0^{B\pi}/F_0^{BK} \simeq 0.9$ in both LCSR and BSW models,
and light quark masses enter through $R_4 = 2{m_K^2/(m_b-m_u)(m_s+m_u)}$.
Although at present \cite{jima} $K^{+} \pi^0/ K^{+} \pi^- \approx 1$
seems to favor \cite{DHHP} large $m_s$
to suppress the penguin $a_6$ term,
for more sensible $m_s < 200$ MeV values,
$K^+\pi^0$ is always visibly less than $K^+\pi^-$ \cite{HHY},
as can be seen in Fig. 2(a).

For $K^{*+} \pi^0/ K^{*+} \pi^-$, the $a_6$ term is absent,
but the $a_2$ and EWP $a_9$ terms are modulated by the factor
$r_1 = f_\pi A_0^{BK^*} /f_{K^*} F_1^{B\pi} =$ 0.9 (0.6)
in LCSR (BSW) model, and
\begin{eqnarray}
\frac{K^{*+}\pi^0}{K^{*+}\pi^-}
 \approx {1\over 2} \left\vert 1 + r_1 \,
{\lambda \vert\frac{V_{ub}}{V_{cb}}\vert e^{-i\gamma}\, a_2
                    + {3\over2} a_9
      \over \lambda \vert\frac{V_{ub}}{V_{cb}}\vert e^{-i\gamma}\,  a_1
                               +  a_4}
\right\vert^2 \approx 0.7\ (0.6)\ {\rm in\ LCSR\ (BSW)},
\end{eqnarray}
as can be seen from Fig. 4(a).

For $K^{*+} \rho^0/ K^{*+} \rho^-$,
$r_1$ is replaced by a more complicated
ratio of $\rho$ and $K^*$ decay constants and $B\to V$ form factors,
and
\begin{eqnarray}
\frac{K^{*+} \rho^0}{K^{*+} \rho^-}
 \approx {1\over 2}\left\vert 1 + r_2 \,
 {\lambda \vert\frac{V_{ub}}{V_{cb}}\vert e^{-i\gamma}\,  a_2
                     + {3\over2} a_9
      \over \lambda\vert\frac{V_{ub}}{V_{cb}}\vert e^{-i\gamma}\, a_1
                              +  a_4} \right\vert^2
 \approx 1,
\end{eqnarray}
since $r_2\simeq 1.2$ turns out to be larger than $r_1$.

\begin{figure}[htb]
\vspace{1cm}
 \leftline{\epsfig{figure=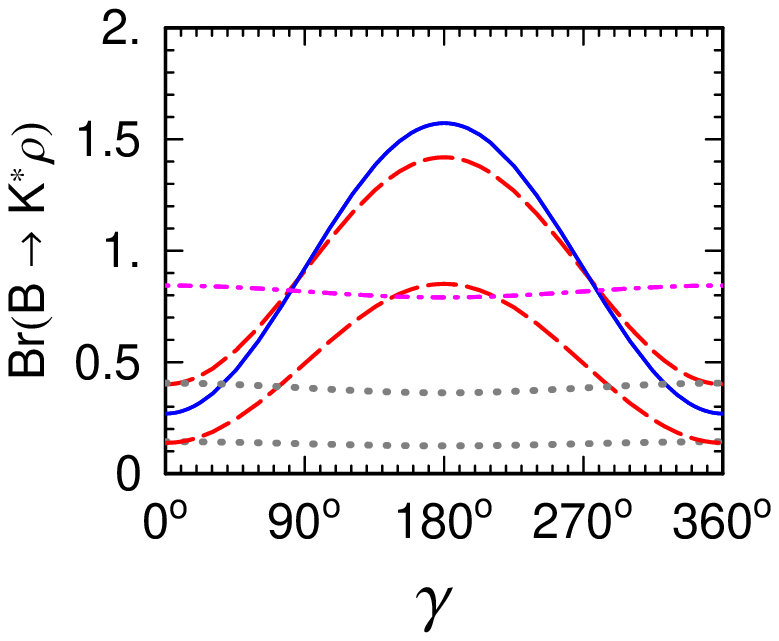,width=7.7cm,height=6cm}
 ~~~~~~ \epsfig{figure=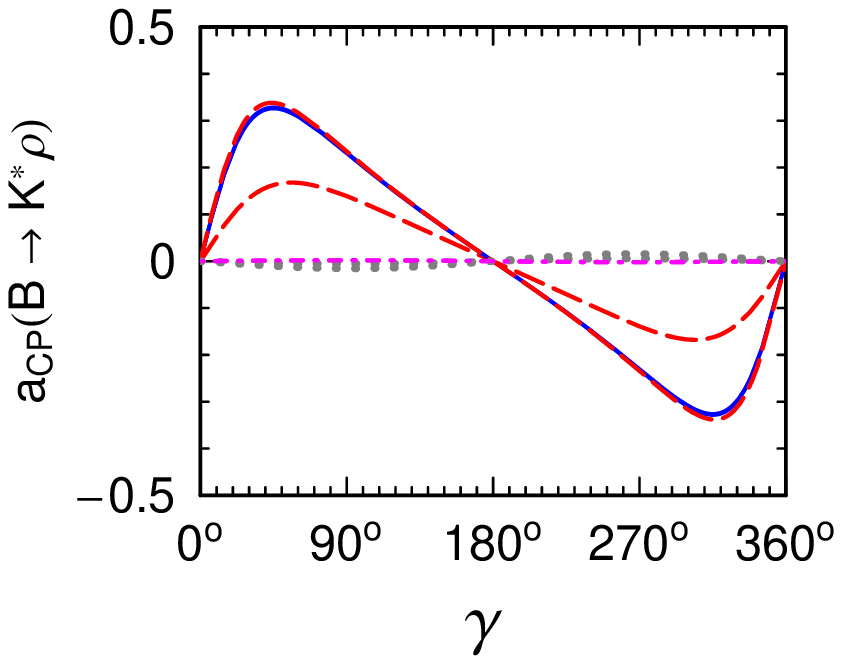,width=7.7cm,height=6cm}}
\vspace{0.8cm}
 \caption {$Br$s and $a_{CP}$s vs. $\gamma$ where solid,
 dash, dotdash and dots are for
 $K^{*+}\rho^-$, $K^{*+}\rho^0$, $K^{*0}\rho^+$ and $K^{*0}\rho^0$,
 respectively.
 Setting the EWP term $a_9 = 0$ lowers (raises) the
 $K^{*+}\rho^0$ ($K^{*0}\rho^0$) rate,
 while the upper $a_{CP}$ curve for $K^{*+}\rho^0$ becomes
 very close to the $K^{*+}\rho^-$ case.
\label{fig:Krho}} \vspace{0.5cm}
\end{figure}

Thus, {\it the EWP effect is most prominent in the $K^{*}\rho^0$ modes},
which enhances the ratio $K^{*+} \rho^0/ K^{*+} \rho^-$ to be close to 1.
It also suppresses the $K^{*0}\rho^0$ mode.
To illustrate this we show in Fig.~\ref{fig:Krho} both the cases
of keeping $a_9$ and with $a_9$ set to 0.
Thus, we see that {\it the EWP effect is able to
enhance the $K^{*+}\rho^0$ rate by a factor of 2!}
In comparison, the EWP effect in $K^+\pi^0$ is
diluted by the additional strong penguin contribution from $a_6$,
while for $K^{*+} \pi^0/ K^{*+} \pi^-$,
it is subdued by the form factor ratio $r_1$.
If $r_1$ is even larger than LCSR case,
then $K^{*+} \pi^0/ K^{*+} \pi^-$ could be closer to 1.
We note that the rate difference between $K^* \rho^0$ and $K^* \omega^0$
(which we discuss below) modes is also mainly due to the EWP contribution.

We find that, for $\gamma\sim 60^\circ-70^\circ$ one has
$K^{*0}\rho^+ \gtrsim K^{*+}\rho^0 \approx K^{*+}\rho^- \gg K^{*0}\rho^0$,
but for $\cos\gamma < 0$ it becomes
$K^{*+} \rho^- \gtrsim K^{*+} \rho^0 > K^{*0} \rho^+ \gg K^{*0} \rho^0$.
The $a_{CP}$s of $K^{*+}\rho^-$ and $K^{*+}\rho^0$ modes
are sizable and have opposite sign to
$\rho^+\rho^-$ and $\rho^+\omega^0$ modes.
For $\gamma\sim 65^\circ$ they could be as large as
$30\%$ and 18\% respectively,
but are of order 15\% or 10\% for $\gamma\sim 120^\circ$.

\subsection{$B \to K^*\omega$ and $K^*\phi$ Modes}

\begin{figure}[htb]
\vspace{1cm} \leftline{\epsfig{figure=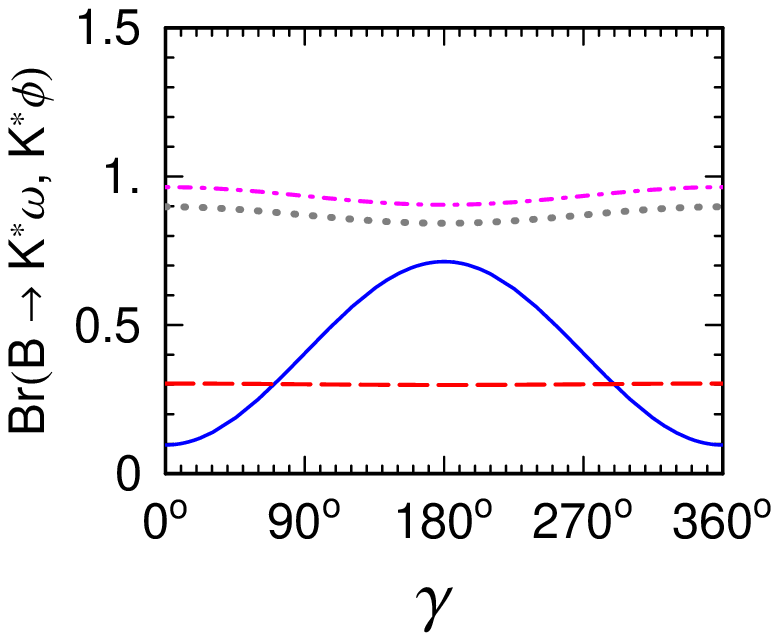,width=7.7cm}
    \ \epsfig{figure=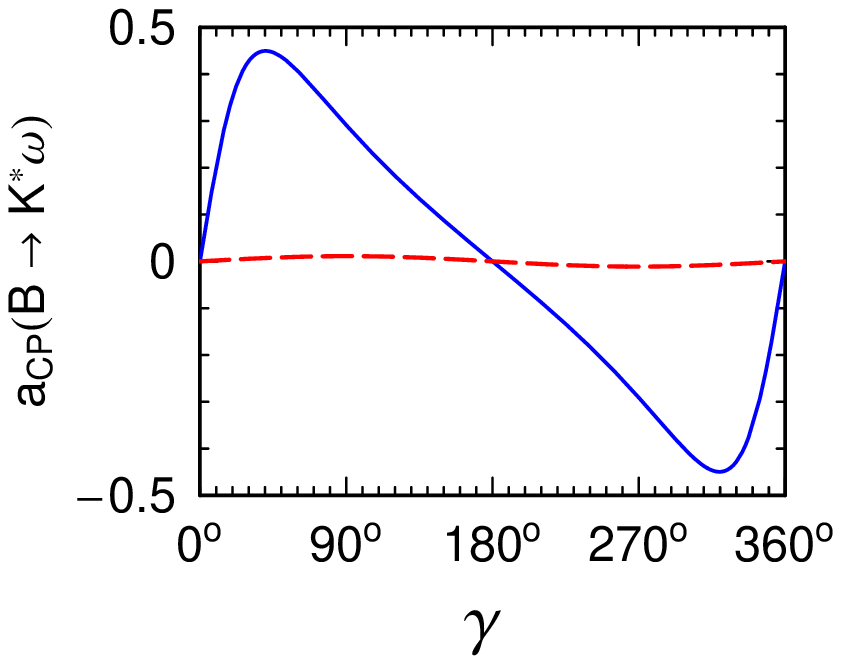,width=8.2cm}}
\vspace{0.8cm}
 \caption {$Br$s and $a_{CP}$s vs. $\gamma$
 where solid, dash, dotdash and dots are for
 $K^{*+}\omega^0$, $K^{*0}\omega^0$, $K^{*+}\phi^0$ and $K^{*0}\phi^0$,
 respectively. The $a_{CP}$s of $K^* \phi$, not shown here, are
 consistent with zero.
\label{fig:Komega}} \vspace{0.5cm}
\end{figure}

The sign of $T$--$P$ interference in $K^{*+} \omega^0$ and
$K^{*+}\rho^0$ modes are rather similar under factorization.
Thus, the $K^{*+} \omega^0$ rates are
also enhanced in the region of $\cos\gamma<0$,
as can be seen in Fig.~\ref{fig:Komega}.
The $K^{*0}\omega^0$ rate is insensitive to $\gamma$
because its tree contribution is color suppressed.
Thus, the $K^{*+} \omega^0$ rate can be
$1.5-2.5$ times larger than $K^{*0}\omega^0$ for $\cos\gamma < 0$,
while $K^{*+} \omega^0 \lesssim K^{*0}\omega^0$ for
$\gamma\sim 60^\circ-70^\circ$.
Since $T/P$ is of order $20\%-30\%$,
direct $a_{CP}$ for $K^{*+} \omega^0$ could
reach 40\% for $\gamma\sim 60^\circ - 70^\circ$,
and could still be 20\% even for $\gamma\sim 120^\circ$.

The $K^* \phi^0$ modes arise from
the pure penguin $b\to s \bar ss$ process
and have very weak $\gamma$ dependence (Fig.~\ref{fig:Komega}).
Though not useful for extracting $\gamma$, they give a more direct
test of the factorization hypothesis.
In the standard model the $a_{CP}$s are
practically zero and any measurement $\geq 10\%$ would
likely be an indication for new physics~\cite{HHY98}.

\subsection{$B \to K^*\eta$ modes}

As pointed out in Ref. \cite{HHY},
having $\cos\gamma < 0$ could explain
the observed splitting of $K^+\eta^\prime > K^0\eta^\prime$,
although the $K\eta^\prime$ modes seem to have
a large singlet contribution, such as coming from the anomaly \cite{HT}.
Even assuming $N_{\rm eff}(LL) = 2 \neq N_{\rm eff}(LR) = 5$ \cite{CCTY}
and low $m_s$ values, the rates
fall $30\%-40\%$ short of observed, while for $N_{\rm eff} = 3$
one can only account for less than half the observed rate.
Since we do not know how to take the anomaly effect into proper account
for exclusive modes, we shall not plot the results here.

\begin{figure}[htb]
\vspace{1cm}
    \leftline{\epsfig{figure=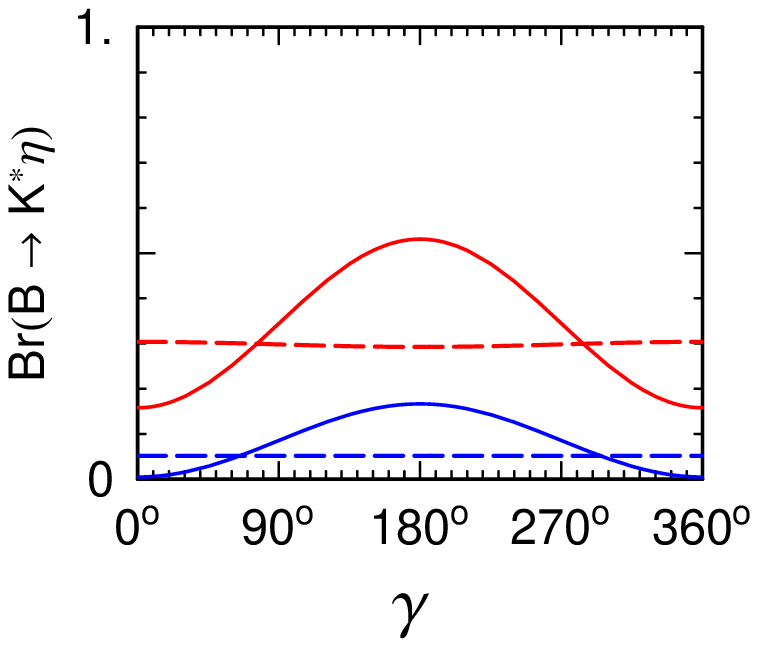,width=7.7cm,height=6cm}
    \ \epsfig{figure=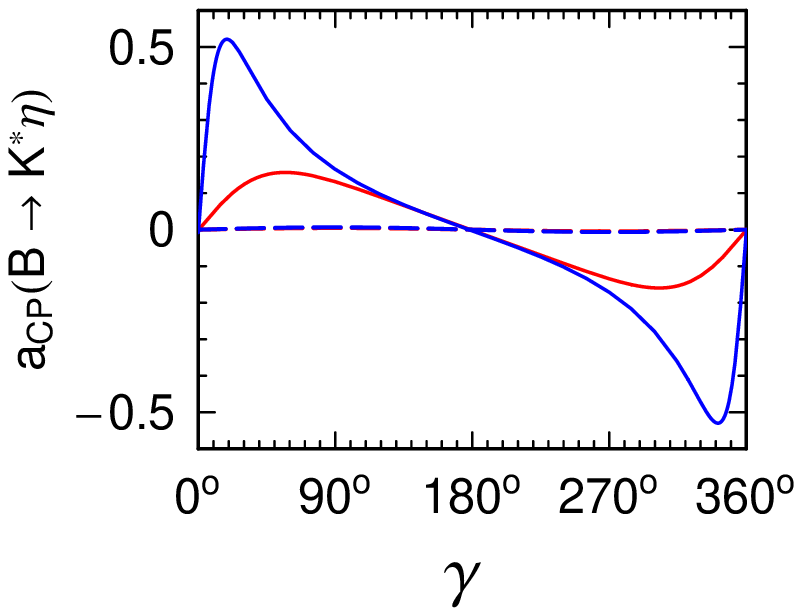,width=7.7cm,height=6cm}}
\vspace{0.8cm}
 \caption {$Br$s and $a_{CP}$s vs. $\gamma$
  where solid and dash are for
 $K^{*+}\eta$ and $K^{*0}\eta$, respectively, for $m_s = 105$ and $200$ MeV.
 Upper curves for $Br$s are for $m_s= $ 105 MeV while for $a_{CP}$s the
 sharper curve is for $m_s = $ 200~MeV.
\label{fig:Keta} } \vspace{0.5cm}
\end{figure}

The $K^*\eta$ modes, however, should be
less susceptible to the anomaly effect,
and with T/P structure similar to $K^* \pi^0$~\cite{HHY}.
Ignoring the extra anomaly term and omitting an overall factor of
$\sqrt{2} G_F m_{K^*}\, \epsilon_{K^*}\cdot p_B$,
one has
\begin{eqnarray}
{\cal M}_{K^{*0}\eta^{(\prime)}}
& \cong & V_{us}^*V_{ub}\, f^u_{\eta^{(\prime)}} A_0\, a_2
        - V_{ts}^*V_{tb}\, \left[\left(f_{K^*} F_1
                                      +f_{\eta^{(\prime)}}^s A_0\right)\, a_4
  \right.\nonumber \\
&& {\hskip 4.15cm}  \left.
                   - \left(f_{\eta^{(\prime)}}^u
                          -f_{\eta^{(\prime)}}^s\right) A_0
                         \left(a_6 Q^{(\prime)} - {1\over 2}a_9\right)
               \right],
  \nonumber \\
{\cal M}_{K^{*+}\eta^{(\prime)}}
& \cong & {\cal M}_{K^{*0}\eta^{(\prime)}}
        + V_{us}^*V_{ub}\, f_{K*} F_1 a_1\,,
\end{eqnarray}
where $Q^{(\prime)}=-m^2_{\eta^{(\prime)}}/(m_b+m_s)m_s$,
$F_1=F^{B\eta^{(\prime)}}_1(m_{K^*}^2)$,
$A_0=A^{BK^*}_0(m_{\eta^{(\prime)}}^2)$ and we have dropped terms that are
much smaller than those shown. Numerically we use $f^u_\eta,\
f^s_\eta=$78, $-112$ MeV, and $f^u_{\eta^\prime},\
f^s_{\eta^\prime}=$63, 137 MeV \cite{CCTY}. The $\gamma$
dependence for $K^{*0}\eta$ mode is weak because the tree
contribution is color suppressed. For $K^{*+}\eta$ one has
constructive $T$--$P$ interference for $\cos\gamma < 0$ hence
$K^{*+}\eta > K^{*0}\eta$ while $K^{*+}\eta \lesssim K^{*0}\eta$
for $\gamma\sim 60^\circ-70^\circ$. As shown in
Fig.~\ref{fig:Keta}, the rates depend strongly on $m_s$, the
strange quark mass. We find $K^{*+}\eta / K^{*0}\eta\approx 1.5$
for $m_s=105$~MeV, but may be enhanced to 2.2 for $m_s=200$~MeV.
The rates could be larger by 50\% or more since $A_0^{BV}$
seems to be larger \cite{HHY} than $F_1^{BP}$, as indicated by the
strength of the $\rho^0\pi^+$ mode.

\subsection{Various Suppressed Modes\label{sec:supp}}

The $\rho^0 \rho^0, \rho^0 \omega^0$, and $\omega^0 \omega^0$ modes
are color suppressed and dominated by penguin contributions
which have opposite sign compared to
$\rho^+\rho^{-,0}$ and $\rho^+ \omega^0$ case.
The rates are enhanced for ${\rm cos}\gamma<0$
but are, however, only of order $10^{-7}$.
The $\rho^0 \phi^0$, $\rho^+ \phi^0$ and $\omega^0 \phi^0$ modes
are pure penguin processes with amplitudes
$\propto V_{td}^* V_{tb} [a_3+a_5-(a_7+a_9)/2]$.
Their rates are     
too small ($\sim 10^{-8}$) to be measurable soon,
and their $a_{CP}$s are practically zero.

\begin{figure}[htb]
\vspace{1cm}
 \leftline{\epsfig{figure=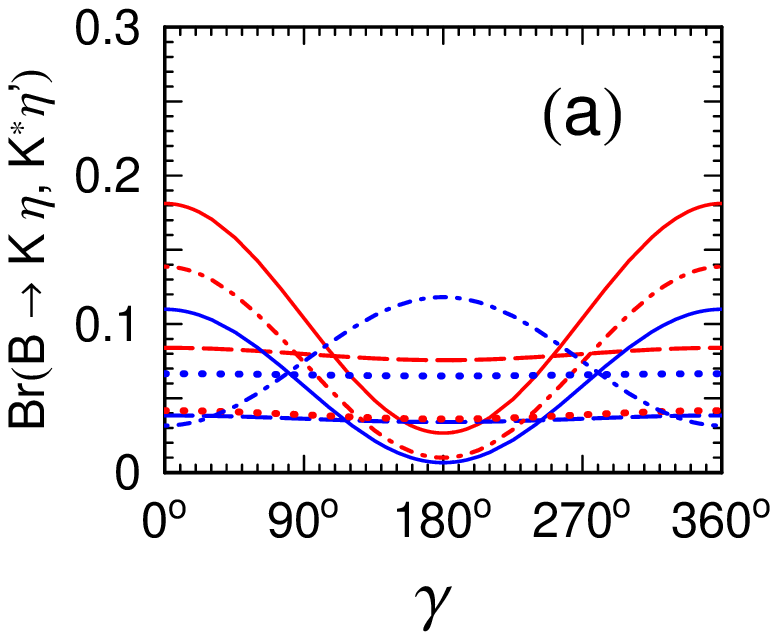,width=7.7cm,height=6cm}
    \ \epsfig{figure=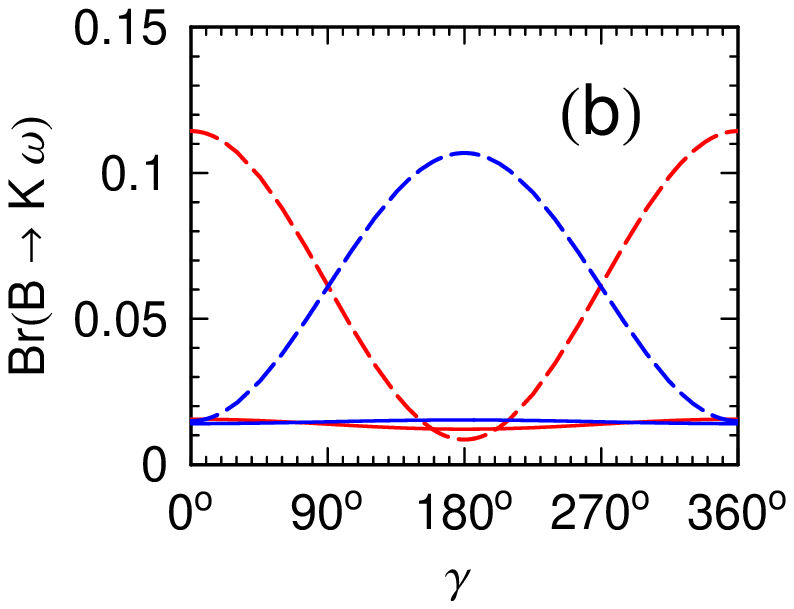,width=7.7cm,height=6cm}}
\vspace{0.8cm}
 \caption {$Br$s vs. $\gamma$ where
 (a) solid, dash, dotdash and dots are for
 $K^{+}\eta$, $K^{0}\eta$, $K^{*+}\eta^\prime$ and $K^{*0}\eta^\prime$,
 respectively;
 (b) solid and dash are for $K^{0}\omega^0$ and $K^{+}\omega^0$,
 respectively.
 The upper (lower) curves for $K\eta$ and $K^{0}\omega^0$
 at $\gamma=180^\circ$ are for $m_s=$105 (200)~MeV,
 while for $K^*\eta^\prime$ and $K^{+}\omega^0$ it is the reverse.
\label{fig:vpKomega} } \vspace{0.5cm}
\end{figure}

The $K^* \eta^\prime$ modes are suppressed because
$f^s_{\eta^\prime} > 0$, as can be seen from Eq. (4.4).
Likewise, $K\eta$ modes are also suppressed.
The $Br$s are given in Fig.~\ref{fig:vpKomega}.
We see that $K^* \eta^\prime \lesssim 1.5\times 10^{-6}$,
and for $\cos\gamma < 0$ the $K^+\eta$ rate is suppressed,
leading to $K^+\eta \lesssim K^0\eta \lesssim 10^{-6}$.
These suppressed modes should be compared with
the $K\eta^\prime$ modes, which are already observed and
are the largest exclusive rare hadronic decays,
and the $K^*\eta$ modes,
which have some chance of being observed in the near future.

The $K^+\omega^0$ mode is reported at the rather sizable level of
$1.5 \times 10^{-5}$ \cite{omegaK}, in strong conflict with the
rather suppressed factorization expectation [See Note Added.].
This is also illustrated in Fig.~\ref{fig:vpKomega} together with
$K^0\omega^0$, which has lower reconstruction efficiency.
The $K\omega^0$ rates are also very sensitive to $m_s$,
but we do not see any way to enhance them within factorization approach.

In general, when modes are suppressed because of
cancellation of different contributions such as
the modes shown in Fig.~\ref{fig:vpKomega},
one is not only sensitive to form
factors and long distance effects, but also sensitive to actual
values of short distance coefficients.

\section{Discussion and Conclusion}\label{dc}

The $B\to VV$ decay rates are quite sensitive to the chosen form
factor model, but the relative sizes of $Br$s and $a_{CP}$s are
not. All $Br$s could easily be larger by 50\% or more if $B\to V$
form factors are in general larger \cite{HHY} than $B\to P$ form
factors, as indicated by the strength of the $\rho^0\pi^+$ mode
\cite{gaofrank}. Our main results are insensitive to $N_{\rm
eff}\neq 3$. For $N_{\rm eff}<3$, the $\rho^+ \rho^0$ and
$\rho^+\omega^0$ modes are enhanced and become closer to
$\rho^+\rho^-$.
For $N_{\rm eff}=2$, the $\rho^0 \rho^0$ and $\omega^0\omega^0$ modes
become one order of magnitude larger, but still below $10^{-6}$.

Subsequent to Ref. \cite{HHY},
the observation of $\rho^\pm\pi^\mp$ and $K^{*+}\pi^-$ modes
\cite{gaofrank} were reported, which offer
further support for the factorization and $\cos\gamma < 0$ hypotheses.
We believe that the
$\pi^+\pi^-$, $\pi^+\pi^0$,
$\rho^+\pi^0$, $\omega^0\pi^+$,
$K^{*+}\pi^0$ and $K^{*0}\pi^+$ modes, all discussed in Ref. \cite{HHY},
would likely emerge with full CLEO II and II.V datasets [See Note Added.].
The $K^{(*)0}\pi^0$ modes are borderline,
the $\rho K$ modes unlikely,
while $\pi^0\pi^0$ and $\rho^0\pi^0$ modes
should not be seen soon if factorization is correct.
But what are the modes discussed here that are promising for detection
in the near future?
As mentioned in Ref. \cite{HHY},
the theoretical computation of VV and $\eta^{(\prime)}$ modes
are less trustworthy even under factorization assumption,
as they depend on vector form factors
or $\eta^{(\prime)}$ decay constants.
We give a discussion nevertheless.

Since helicity angle methods (boosted $\pi^+$, $K^+$ or $\pi^0$
along parent $\rho^{+,0}$ or $K^{*+,0}$ momentum)
seem promising from observed $\rho^\pm\pi^\mp$ reconstruction \cite{gaofrank},
the modes $K^{*+}\rho^0$ and $\rho^+\rho^0$ with $\rho^0\to \pi^+\pi^-$
can probably be reconstructed above background.
It is less clear whether this is the case for
$\rho^+\rho^-$ and $K^{*+}\rho^-$.
The $K^{*0}\rho^+$ mode is at best borderline
even without considering background,
while $K^{*0}\rho^0 \sim 10^{-6}$ is unlikely to be observed soon.

The reconstruction of two body modes containing an $\omega^0$
has been shown \cite{omegaK} to be of low background
and with efficiency better than $\eta^\prime$ modes.
Assuming that the $B\to V$ form factors $A_{1,2}$ and $V$
are similarly enhanced as $A_0$,
the $\rho^+\omega^0$ and perhaps the $K^{*+}\omega^0$ modes
could be observed soon, while $K^{*0}\omega^0$ is at best borderline.
The four $K^{(*)}\phi^0$ modes should also suffer little from background.
The $K^*\phi^0$ modes could be split above $K\phi^0$ modes
if $B\to K^*$ form factors are enhanced over $B\to K$.
At the $0.5\times 10^{-5}$ level,
the $K^+\phi^0$ and $K^{*+}\phi^0$ modes are likely to appear soon,
while $K^0\phi^0$ and $K^{*0}\phi^0$ modes
suffer from detection efficiency and may be borderline.
The $K^{*+}\eta$ mode could emerge if $A_0^{BK^*}$ is large,
but $K^{*0}\eta$ is probably borderline.
These modes should again have low background.

All the suppressed modes mentioned in Sec.~\ref{sec:supp} should
not appear. The nonobservation of $K\bar K$ modes so far suggest
inelastic final state rescattering effects are small.
However, the observation of $ K\omega$, $K\rho$ or
any of the suppressed modes under factorization may indicate
the size of final state rescattering,
hence the level of breakdown of factorization.
We cannot account for the observed large $K^+\omega^0 \sim 1.5 \times
10^{-5}$ in factorization approach, and await further updates with
full CLEO II and II.V datasets [See Note Added.].

In conclusion, we have studied the $\gamma$
dependence of hadronic rare B decays to PP, PV, VV and $K^*\eta$ modes
within the factorization approach.
We find that light cone sum rule form factors give better fit to
$B\to PP,$ $PV$ data.
The $\rho^+\omega^0$, $\rho^+\rho^0$, $K^{*+}\rho^0$
and $K^{(*)+}\phi^0$,
and perhaps the $K^{*+}\omega^0$ and $K^{*+}\eta$ modes,
should be observable with the full CLEO II and II.V datasets.
Whether the sizable $\rho^+\rho^-$ and $K^{*+}\rho^-$ modes can be observed
depends crucially on the background level,
while the clean modes of $K^{(*)0}\phi^0$
are probably borderline because of statistics.
The $K^{*0}\rho^0$, $K^{*0}\omega^0$ and $K^{*0}\eta$ modes
are likely too low to be seen with $10^7$ $B\bar B$s.
Finding $\rho^+\omega^0 < \rho^+\rho^0$,
$K^{*+}\rho^{-,0} > K^{*0}\rho^+$,
$K^{*+}\omega^0 > K^{*0}\omega^0$
and $K^{*+}\eta > K^{*0}\eta$ would support $\cos\gamma < 0$.
The EWP effect should be most prominent in
$K^{*+}\rho^0$ mode as compared to $K^{(*)+}\pi^0$,
leading to a factor of two enhancement in rate,
and observation of $K^{*+}\rho^0 \simeq K^{*+}\rho^-$
would give strong evidence for the electroweak penguin.
The weakly $\gamma-$dependent pure penguin processes $K^{(*)}\phi^0$
can be used as a direct test of the factorization hypothesis.
If large $a_{CP}$ is measured in $K^{(*)+}\phi^0$ modes,
then new physics would be implied.

The rare $B \to VV$ modes should also be studied with vigor!

\vskip 0.7cm \centerline{\bf Acknowledgement}
\vskip 0.15cm
This work is supported in part by
the National Science Council of R.O.C.
under Grants NSC-88-2112-M-002-033 
and NSC-88-2112-M-001-006.
We thank X.G. He for many useful comments,
and B. Behrens, J.G. Smith and F. W\" urthwein for discussions.

\vskip 0.7cm
\noindent{\bf Note Added.}
\vskip 0.15cm

At the completion of this paper,
CLEO announced \cite{hf8} new results at the
``{\it 8}th International Symposium on Heavy Flavor Physics"
held at Southampton, England.
The long sought-after $\pi^+\pi^-$ mode is found at
$(0.47^{+0.18}_{-0.15}\pm 0.13)\times 10^{-5}$.
The $\omega K^+$ mode has disappeared under the 90\% upper limit
of $<0.8\times 10^{-5}$, in strong contrast to the
published value of $(1.5^{+0.7}_{-0.6}\pm 0.2)\times 10^{-5}$ \cite{omegaK}.
At the same time, the previously unmeasured
$\omega\pi^+$ mode is now measured at
$(1.1\pm 0.3\pm 0.1)\times 10^{-5}$.
The $K^+\pi^-$ mode is also updated to
$(1.88^{+0.28}_{-0.26}\pm 0.06)\times 10^{-5}$
and now larger than $K^+\pi^0$.
All these new results are in better agreement with
the discussions presented in this paper.
There is no indication of breakdown of factorization
in rare B decays so far,
so long that one takes $\cos\gamma < 0$.

\newpage

\vspace{1cm}

\end{document}